\documentclass[twoside]{article}
\usepackage{fleqn,espcrc2}
\usepackage{epsf}
\usepackage{pstcol,pst-node,pst-coil}
\usepackage{graphics}
\usepackage[figuresright]{rotating}

\newcommand{\nn}{\nonumber}

\def\vev#1{\left\langle #1 \right\rangle}
\def\abs#1{\left|{\bf #1}\right|}

\def\Dsl{\hbox{/\kern-.6000em D}} 
\def\vabsq#1{\left|{\bf #1}\right|^2}
\def\vabs#1{\left|{\bf #1}\right|}
\def\dsl{\,\raise.15ex\hbox{/}\mkern-13.5mu D}
\def\bsigma{\mbox{\boldmath $\sigma$}}
\def\lqcd{\Lambda_{\rm QCD}}

\def\bsigma{\mbox{\boldmath $\sigma$}}
\def\OMIT#1{}

\def\lesssim{\ \raise.3ex\hbox{$<$\kern-.75em\lower1ex\hbox{$\sim$}}\ }

\title{Nonrelativistic Bound States in Quantum Field Theory%
\thanks{UCSD/PTH 00-27. Talks presented at Lattice 2000 and SPIN 2000}
}
\author{Aneesh V. Manohar\address{Department of Physics,
University of California at San Diego,\\
9500 Gilman Drive, La Jolla, CA 92093-0319} and
Iain W. Stewart\addressmark}

\begin{document}

\begin{abstract} 
Nonrelativistic bound states are studied using an effective field theory. Large
logarithms in the effective theory can be summed using the velocity
renormalization group. For QED, one can determine the structure of the leading
and next-to-leading order series for the energy, and compute corrections up to
order $\alpha^8 \ln^3 \alpha$, which are relevant for the present comparison
between theory and experiment. For QCD, one can compute the velocity
renormalization group improved quark potentials. Using these to compute 
the renormalization group improved $\bar t t$ production cross-section near
threshold gives a result with scale uncertainties of 2\%, a factor of 10
smaller than existing fixed order calculations.
\vspace{1pc}
\end{abstract}

\maketitle

\section{INTRODUCTION}

Nonrelativistic bound states in QED and QCD provide an interesting and highly
nontrivial problem to which effective field theory methods can be
applied~\cite{Caswell,BBL}. The QCD bound states we will consider are heavy
$\bar Q Q$ states such as $\bar tt$ bound states or the $\Upsilon$ system. In
QED, the classic examples are Hydrogen, muonium ($\mu^+e^-$), and positronium.
Each of these systems has three important scales, $m$ the fermion mass, $mv$
the fermion momentum, and $mv^2$, the fermion energy. (For Hydrogen and
muonium, $m$ is the electron mass or the reduced mass of the two particles.)
The velocity $v$ is of order the coupling constant ($\alpha_s$ or $\alpha$),
and we will only consider the case $v\ll1$, $mv^2\gg\lqcd$ so that
nonperturbative effects are small. 

Multiscale problems with widely separated scales are well suited for study
using effective field theories. For example, if the problem has the scales $m_1
\gg m_2 \gg m_3 \ldots$, one first starts with the theory above $m_1$, and
matches to an effective theory below $m_1$ in which only modes with masses much
smaller than $m_1$ are retained. The effective theory is then scaled using the
renormalization group to the next scale $m_2$. At this point, particles with
mass $m_2$ are integrated out to construct a new effective theory, and so on.
The complicated multiscale computations of the original theory are reduced to a
number of simpler single scale computations of matching and running in the
effective theory.  The effective theory method also allows one to sum
logarithms of the ratio of mass scales $\ln m_i/m_{i+1}$ using the
renormalization group evolution between $m_i$ and $m_{i+1}$.

The goal is to correctly separate the scale $m$, $mv$ and $mv^2$ for
nonrelativistic bound state problems using an effective field theory, and to
sum large logarithms using the renormalization group. The large logarithms in
this case are $\ln p/m$, $\ln E/m$ and $\ln p/E$ which are proportional to $\ln
v$, and lead to $\ln \alpha$ contributions to bound state energies.
Furthermore, for QCD, the effective theory also determines the scale of the
strong coupling constant, i.e.\ whether one should use $\alpha_s(m)$,
$\alpha_s(mv)$ or $\alpha_s(mv^2)$. The nonrelativistic effective theory,
NRQCD/NRQED, has been studied extensively in the 
past~\cite{Caswell,BBL,Labelle,LM,AM,GR,LukeSavage,PS1,PS2,PS3,LMR,amis,%
amis2,amis3,amis4,Brambilla,Kniehl}. What is new is the
precise formulation of the effective theory, and the way in which the
renormalization group is scaling is implemented.

The results presented here will be applied to the study of $\bar tt$ production
in the threshold region. There is a large ratio of scales, $m_t \sim 175$~GeV,
$m_t v \sim 26$~GeV and $m_t v^2 \sim 4$~GeV, where $v \sim 0.15$ is the typical
velocity in the nonrelativistic bound state. Clearly $\alpha_s \ln v$ is not
small, and summing logarithms is important in this case.

The results are also useful in QED. While $\alpha \ln \alpha$ is small, it is
important to compute to high orders because the experiments have high
precision. The Hydrogen Lamb shift of 1057.845~MHz is known to an accuracy of
9~KHz~\cite{Lundeen}, the Hydrogen hyperfine splitting is measured to be
\hbox{1420.405 751 766 7(9)}~MHz~\cite{Hellwig}, and the muonium hyperfine
splitting is \hbox{4463.302 776(55)}~MHz~\cite{Liu}. The binding energy of
Hydrogen, $m_e \alpha^2/2$ is $2 \times 10^{10}$~MHz, so the experimental error
in the Lamb shift of 10 ppm is a part in $10^{12}$ of the binding energy. We
will be able to compute corrections of order $m_e \alpha^8 \ln^3
\alpha/(4\pi)^2 \sim 5$~KHz to the Lamb shift, which are relevant for the
present comparison between theory and experiment. [The counting of $4\pi$
factors for bound states is a little different than the conventional
counting~\cite{GM}. Potential loops give powers of $\alpha$ whereas soft and
ultrasoft loops give powers of $\alpha/(4\pi)$.]

A detailed comparison of theory and experiment for QED can be found in
Refs.~\cite{kinoshita,pachuki,eides}.

\section{NEW RESULTS}

There are many interesting new results that have been obtained for QED and
QCD~\cite{LMR,amis,amis2,amis3,amis4,mss1,amis5,hmst1,hmst2}. For QED, one
finds a universal description of $\ln \alpha$ terms. A single renormalization
group equation gives the Lamb shift, hyperfine splitting and decay widths for
Hydrogen, muonium and positronium. The renormalization group method allows us
to compute for the first time the $\alpha^8 \ln^3\alpha$ Lamb shift in
positronium and the $\alpha^8 \ln^3\alpha$ Lamb shift in Hydrogen and muonium
including recoil corrections. It also resolves a controversy in the literature
about the $\alpha^8 \ln^3\alpha$ Hydrogen Lamb shift in the limit $m_p \to
\infty$.

The renormalization group method allows one to understand the structure of the
QED perturbation series, and why the $\ln \alpha$ corrections terminate. The
leading order series has a single term that contributes at order $\alpha^5 \ln
\alpha$ to the energy, and the next-to-leading order series terminates after
three terms,  $\alpha^6 \ln \alpha$, $\alpha^7 \ln^2 \alpha$, and $\alpha^8
\ln^3 \alpha$. One also finds some infinite series of terms in QED, but they
have the form $(\alpha^3 \ln^2 \alpha)^n$, rather than $(\alpha \ln \alpha)^n$.

In QCD, one is able to sort out the scales for $\alpha_s$, and decide whether
the strong coupling is $\alpha_s(m)$, $\alpha_s(mv)$, or $\alpha_s(mv^2)$. We
also obtain the renormalization group improved computations of the bound states
potentials in QCD. There are numerous applications of these results, and I will
show an example of the dramatic improvement one obtains for the $\bar t t$
production cross-section near threshold~\cite{hmst1,hmst2}.

\section{THE PROBLEM}

\begin{figure}
  \centerline{\epsfxsize=3truecm \epsfbox{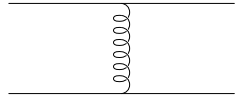}}
\caption{A potential gauge boson exchange. The typical momentum and energy 
transfered are $mv$ and $mv^2$.}
\label{fig:5a}
\end{figure}

\begin{figure}
  \centerline{\epsfxsize=6truecm \epsfbox{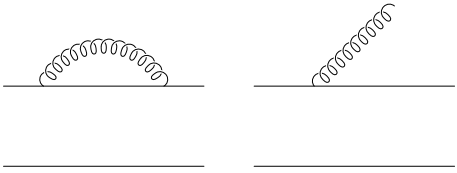}}
\caption{Graphs containing ultrasoft photons, with energy and momentum of order
$mv^2$.}
\label{fig:5b}
\end{figure}

\begin{figure}
  \centerline{\epsfxsize=6truecm \epsfbox{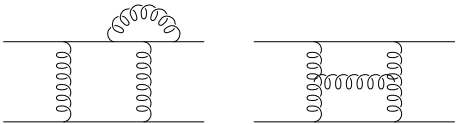}}
\caption{Graphs containing gauge bosons carrying momentum of order $mv$ and
$mv^2$.}
\label{fig:5c}
\end{figure}

\begin{figure}
 \centerline{\epsfxsize=5truecm \epsfbox{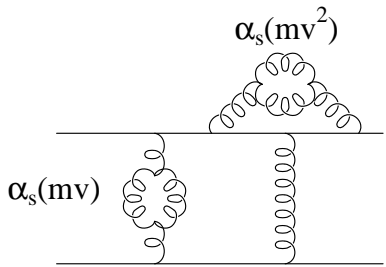}}
\caption{An example of a graph involving both $\alpha(mv)$ and $\alpha(mv^2)$.}
\label{fig:5d}
\end{figure}

The basic problem can be seen by drawing a few Feynman diagrams. A typical
gauge boson exchange in the $t$ channel such as Fig.~\ref{fig:5a} has momentum
transfer of order $p \sim mv$. A wavefunction graph or radiated gauge boson
graph such as Figs.~\ref{fig:5b} have gauge boson momenta of order $E\sim
mv^2$. More interesting diagrams such as those in Figs.~\ref{fig:5c} involve
gauge bosons with momenta of order $p$ and order $E$. In a graph such as
Fig.~\ref{fig:5d}, the vacuum polarization insertions make the effective
coupling of the two gluons $\alpha_s(mv)$ and $\alpha_s(mv^2)$ respectively.
One result which should be clear from Fig.~\ref{fig:5d} is that graphs can
involve $\alpha_s(mv)$ and $\alpha_s(mv^2)$ \emph{simultaneously}. We will
return to this important point later on.

\section{MOMENTUM REGIONS AND DEGREES OF FREEDOM}

The Feynman integrals in the full theory can be evaluated using the threshold
expansion~\cite{beneke}. The important momentum regions (in Feynman gauge) are
referred to in the literature as hard ($E \sim m$, $p \sim m$), potential ($E
\sim mv^2$, $p \sim mv$), ultrasoft ($E \sim mv^2$, $p \sim mv^2$) and soft ($E
\sim mv$, $p \sim mv$). The threshold expansion momentum regions are often used
to describe bound state computations; however it is important to note that
\emph{the threshold expansion is not an effective field theory}. To construct
an effective field theory, one needs to include only modes that can be on-shell.
The effective theory therefore has nonrelativistic fermions (which are
potential modes), and soft and ultrasoft gauge boson modes. The hard fermion
and gauge boson momentum regions, the soft fermion momentum region, and the
potential gauge boson momentum region do not require modes in the effective
theory.

The desired effective theory is valid for energies and momenta much smaller than
the fermion mass $m$. One can try expanding in powers of $E/m$ and $p/m$ as in
heavy quark effective theory, so that the expansion parameter is $1/m$. For
example, the dispersion relation $E=\sqrt{{\bf p}^2+m^2}$ gives terms in the
Lagrangian of the form
\begin{eqnarray}\label{dispersion}
L = \psi^\dagger\left(E-{{\bf p}^2 \over 2m}+{{\bf p}^4 \over 8 m^3}
+\ldots \right)\psi.
\end{eqnarray}
The lowest order propagator is $1/(E+i\epsilon)$, which gives $\theta(t)$ in
position space. This is the static propagator of HQET: fermions propagate
forward in time, but do not move in space. This propagator is acceptable for
some calculations involving heavy quarks. For example, one can compute the
static potential between fixed sources using this propagator. However, for
$\bar t t$ production, the quarks are produced at the same point, and they
remain at the same point for all time if the static propagator is used. This is
too singular, and the HQET expansion breaks down. In general, it is essential
for treating nonrelativistic bound states that the heavy fermions move. For
this to occur, the lowest order propagator should be $1/(E-{\bf
p}^2/2m+i\epsilon)$, so that $E$ and ${\bf p}^2/2m$ are of the same order in
the effective theory power counting. This implies that the $1/m$ expansion
cannot be used; instead one must use an expansion in powers of $v$, where $E$
and ${\bf p}^2/2m$ are both of order $v^2$~\cite{Caswell,BBL}.

The effective theory expansion parameter is the velocity $v$, and formally,
$\alpha$ must also be treated as order $v$. Thus order $\alpha^2$ radiative
corrections to the leading term are just as important as order $v^2$
relativistic corrections. The effective theory below the scale $m$ has:
\begin{itemize}
\item Nonrelativistic fermions with propagator
\[
{1 \over E - {\bf p}^2/2m + i \epsilon}
\]
\item Ultrasoft gauge bosons coupled via interactions that are
multipole expanded~\cite{GR}.
\item Potentials $V({\bf p,p^\prime})$ for the scattering of an incoming $Q$
and $\bar Q$ with momenta ${\bf p}$ and $-{\bf p}$ to outgoing $Q$ and $\bar Q$
with momenta ${\bf p^\prime}$ and $-{\bf p^\prime}$.
\item Soft gauge bosons. The importance of introducing soft fields
in the effective theory was first pointed out by
Griesshammer~\cite{griesshammer}.
\end{itemize}
The effective theory has two different gauge boson fields, soft bosons and
ultrasoft bosons. This does not lead to any double counting if graphs are
evaluated in dimensional regularization.

The static theory is not the $m\to \infty$ limit or the $v \to 0$ limit of the
effective theory. For this reason, the static potential and the effective theory
potential are not equal.

\section{POWER COUNTING}

The power counting parameter of the effective theory is the velocity $v$. If
one expands the dispersion relation as in Eq.~(\ref{dispersion}), then $E$ and
${\bf p}^2/2m$ are both of order $v^2$, and ${\bf p}^4/8m^3$ is of order $v^4$,
i.e.\ of order $v^2$ relative to the leading term. 

The potential $V({\bf p},{\bf p^\prime})$ also has an expansion in powers of
$v$. The leading term is the Coulomb potential, $V({\bf p},{\bf p^\prime})
\propto \alpha/\vabsq k$, where $k={\bf p^\prime}-{\bf p}$ is the momentum
transfer. Since momentum is of order $mv$, the Coulomb potential is naively of
order $\alpha/v^2$. However, the potential is a four-fermion operator, whereas
the kinetic energy is a two-fermion operator. This leads to an additional
factor of $v$ from the power counting factors for the fields, so that the
Coulomb potential is of order $\alpha/v$ in the effective theory. One can then
determine the power counting for all the other potentials by comparing with the
Coulomb potential. The hyperfine interaction $\propto \alpha \mathbf{S}_1 \cdot
\mathbf{S}_2/m^2$ is generated by one-photon exchange, and is of order $v^2$
relative to the Coulomb interaction, so it is of order $\alpha v$ in the power
counting, as are the spin-orbit, tensor and contact (Darwin) interactions. At
one-loop, there are also potentials that are proportional to odd-powers of $\bf
k$. The first such potential is proportional to $\alpha^2/\vabs k$, and is of
order $\alpha^2 v^0$ in the power counting.

\begin{figure}
  \centerline{\epsfxsize=3truecm \epsfbox{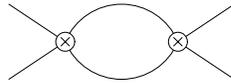}}
\caption{An iteration of two potentials in the effective theory.}
\label{fig:10}
\end{figure}

A loop graph such as Fig.~\ref{fig:10}  of the time-ordered product of two
potentials of order $\alpha^{a_1}v^{b_1}$ and $\alpha^{a_2}v^{b_2}$ is of order
$\alpha^{a_1+a_2}v^{b_1+b_2}$.
One can now see that the static potential differs from the $m\to \infty$ or
$v\to 0$ limit of the effective theory potentials. For example, the loop graph
of Fig.~\ref{fig:10} with one $1/(m \vabs k)$ and one Coulomb potential is of
order $\alpha^2 v^0 \times \alpha/v = \alpha^3/v$, and is of the same order in
$v$ as the Coulomb potential. The two particle intermediate state propagator
$1/(E-{\bf p}^2/2m)=2m/(2mE-{\bf p^2})$ produces a factor of $m$ in the
numerator, that cancels the $1/m$ at the vertex. In the static theory, the 
$1/(m \vabs k)$ potential is set to zero before the loop integration, so that
the graph of  Fig.~\ref{fig:10} is not present in the static theory. As a
result, the  NRQCD potential~\cite{amis5} differs from the static potential.

\section{MATCHING CONDITIONS}

\begin{figure}
  \centerline{\epsfxsize=6truecm \epsfbox{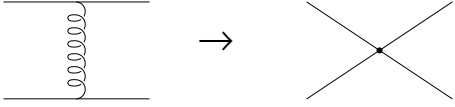}}
\caption{Tree-level matching for the potential.}
\label{fig:13a}
\end{figure}

\begin{figure}
 \centerline{\epsfxsize=6truecm \epsfbox{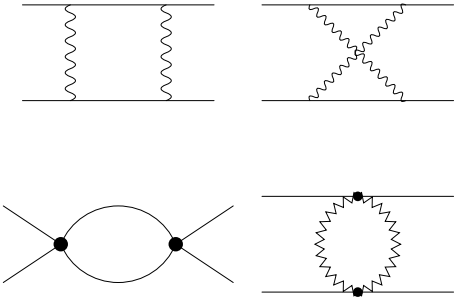}}
\caption{One-loop matching for the potential in QED. The first line gives
examples of the
full theory graphs. The second line gives examples of 
effective theory graphs:
an iteration of two potentials, and a soft photon graph. The difference of the
two sets of graphs gives the one-loop correction to the potential.}
\label{fig:13b}
\end{figure}

\begin{figure}
  \centerline{\epsfxsize=6truecm \epsfbox{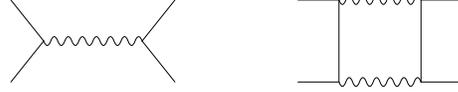}}
\caption{Annihilation contributions to the positronium potentials. The second
graph has an imaginary part. [There is also a one-loop crossed box in the
annihilation channel.]}
\label{fig:13c}
\end{figure}

The method of calculating matching conditions is the same as in any effective
theory. One computes the graphs in the full theory at the scale $\mu=m$, and
subtracts the corresponding graphs in the effective theory. The graph in
Fig.~\ref{fig:13a} gives the matching condition for the fermion potential. The
full theory amplitude,
\begin{equation}
{
\left[\bar u({\bf p^\prime})\gamma^\mu u({\bf p}) \right]
\left[\bar u(-{\bf p^\prime})\gamma_\mu u(-{\bf p}) \right]
\over \left(\mathbf{p-p^\prime}\right)^2}
\end{equation}
is expanded in powers of $\mathbf{p,p^\prime}$, to give the potential in the
effective theory. At one-loop, the difference of the full theory and effective
theory graphs in Fig.~\ref{fig:13b} give the one-loop corrections to the
matching potential. The only difference at this stage between Hydrogen and
positronium is that there are annihilation contributions to the positronium
potential from graphs such as Fig.~\ref{fig:13c}. The graphs can have an
imaginary part, that give the positronium decay width.

\section{RENORMALIZATION GROUP EVOLUTION}

The nonrelativistic bound state system has three important mass scales, $m$,
$mv$ and $mv^2$.

\subsection{Two-stage running}

\begin{figure}
  \centerline{\epsfxsize=4truecm \epsfbox{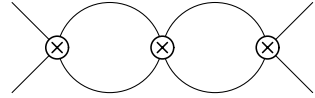}}
\caption{An ultraviolet divergent two-loop graph involving the iteration of
three potentials.}
\label{fig:14a}
\end{figure}

The conventional method of implementing the renormalization group is as follows
\begin{itemize}
\item Start at $\mu=m$
\item Scale $\mu$ from $m$ to $mv$
\item Integrate out the soft modes at $mv$
\item Scale $\mu$ from $mv$ to $mv^2$
\end{itemize}
This is referred to as the two-stage method, because there are two-stages of
renormalization group evolution. Consider a loop graph involving time-ordered
products of potentials, such as Fig.~\ref{fig:14a}.
This graph contains a logarithm of the form $\ln \sqrt{mE}/\mu$. When $\mu$ is
set to $mv$, this logarithm has the from $\ln \sqrt{E/mv^2}$, and is small. Thus
the logarithms in the graph are summed by renormalization group evolution of
$\mu$ from $m$ to $mv$.

\begin{figure}
  \centerline{\epsfxsize=3truecm \epsfbox{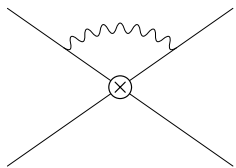}}
\caption{One-loop ultrasoft photon renormalization of
the potential.}
\label{fig:14b}
\end{figure}

The graph in Fig.~\ref{fig:14b} involving an ultrasoft photon exchange contains
a logarithm of the form $\ln E/\mu$.
The $\mu$ in this ultrasoft graph is scaled all the way down (in two stages) to
$mv^2$, at which point the logarithm is $\ln E/mv^2$, and also small.

However, this two-stage method of implementing the renormalization group turns
out to be incorrect for nonrelativistic bound states. The reason is that the
scales $mv$ and $mv^2$ are correlated---one cannot be varied independently of
the other. Instead one needs to use an alternative one-stage scaling procedure.

\subsection{One-stage running}

In one stage running, one introduces two different $\mu$ parameters, $\mu_S$
and $\mu_U$~\cite{LMR}. In dimensional regularization in $4-2\epsilon$
dimensions, the soft photon coupling is multiplied by $\mu_S^{\epsilon}$, the
ultrasoft photon coupling by $\mu_U^{\epsilon}$, and the potentials by 
$\mu_S^{2\epsilon}$. Note that this is only possible because we have two
different photon fields to represent the soft and ultrasoft photons in the
effective theory. Then
\begin{itemize}
\item Set $\mu_S=m\nu$, $\mu_U=m\nu^2$
\item Start at $\nu=1$ and scale to $\nu=v$.
\end{itemize}
This procedure is referred to as the velocity renormalization group, because
one runs in velocity $\nu$ rather than momentum~\cite{LMR}. The logarithms in
Figs.~\ref{fig:14a} and \ref{fig:14b} are now $\ln \sqrt{mE}/m \nu$ and $\ln
E/m\nu^2$, which are minimized when $\nu=v$. Thus this method also minimizes
logarithms in the diagrams, and sums them by renormalization group evolution.

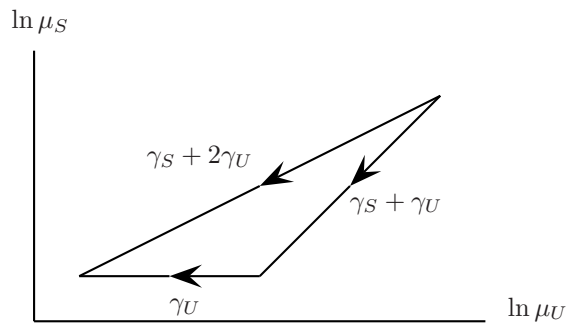
\begin{figure}
\begin{center}
\psset{unit=0.6cm}
\begin{pspicture}(0,0)(10,7.5)
\psline(0,0)(10,0)
\psline(0,0)(0,6)
\psline[arrows=->,arrowscale=2](9,5)(7,3)
\psline(7,3)(5,1)
\psline[arrows=->,arrowscale=2](5,1)(3,1)
\psline(3,1)(1,1)
\psline[arrows=->,arrowscale=2](9,5)(5,3)
\psline(5,3)(1,1)
\put(7,2.5){$\gamma_S+\gamma_U$}
\put(2.5,3.5){$\gamma_S+2\gamma_U$}
\put(3,0.25){$\gamma_U$}
\put(10.5,0.125){$\ln \mu_U$}
\put(-0.5,6.5){$\ln \mu_S$}
\end{pspicture}
 \caption{Paths in the $(\mu_U,\mu_S)$ plane for one-stage and two-stage 
running.}
\label{fig:16}
\end{center}
\end{figure}

The difference between the two renormalization group methods can be seen in
Fig.~\ref{fig:16}~\cite{mss1}.
In two-stage running, there is only a single $\mu$, so that $\mu_S=\mu_U=\mu$,
and they are lowered together from $m$ to $mv$. At this point, the soft modes
are integrated out, and $\mu_U$ for the ultrasoft modes is lowered to $mv^2$.
The integration path in Fig.~\ref{fig:16} is along the lower edges of the
triangle. In one-stage running, the integration path is along the diagonal. It
is convenient to define two anomalous dimensions, $\gamma_S$ and $\gamma_U$ by
taking the derivatives of Green's functions with respect to $\ln \mu_S$ and
$\ln \mu_U$, respectively. One can show by explicit calculation that
\begin{itemize}
\item The two paths give different answers. The integration is path dependent
because $\nabla \times \gamma \not = 0$.
\item One-stage running using the velocity renormalization group agrees with
explicit QED calculations at order $\alpha^3\ln^2\alpha$, $\alpha^7 \ln^2
\alpha$ and $\alpha^8 \ln^3 \alpha$.
\end{itemize}
The moral is that for nonrelativistic bound states, one should run in velocity
rather than momentum.

The difference between the two integration methods can be made more precise. In
the two-stage method, one first integrates $\gamma_S + \gamma_U$ from $\mu=m$
to $\mu=mv$, and then integrates $\gamma_U$ from $\mu=mv$ to $\mu=mv^2$. In the
one-stage method, one integrates $\gamma_S+2\gamma_U$ (since $\ln\mu_U$ runs
twice as fast as $\ln \mu_S$) from $\nu=1$ to $\nu=v$. If the anomalous
dimensions are constant, the two methods give
\begin{eqnarray*}
\begin{array}{ll}
\left(\gamma_S + \gamma_U\right) \ln{mv \over m} + 
\gamma_U \ln{mv^2 \over mv} & \hbox{\qquad two-stage} \\[10pt]
\left(\gamma_S + 2 \gamma_U\right) \ln{v} & \hbox{\qquad one-stage}
\end{array}
\end{eqnarray*}
and agree with each other. However, in general anomalous dimensions are not
constant, but can depend on coupling constants $V_i$, that themselves run. As a
result, one finds that the $\ln v$ terms agree, but the higher order terms
differ. For example, consider a $\ln^2 v$ term that depends on the product of
$\gamma_S$ and $\gamma_U$. For two-stage running, the contribution is proportional to
$\gamma_S \gamma_U + 0 \gamma_U=\gamma_S \gamma_U$ from the two pieces of the
path. For one-stage running, the contribution is $\gamma_S \left(2 \gamma_U
\right)$, which differs by a factor of two. Similarly, a $\gamma_S \gamma_U^2
\ln^3 v$ contribution differs by a factor of four, and so on.

\section{RUNNING POTENTIALS}

The running potential $V(\mathbf{p,p^\prime})$ has an expansion
\begin{equation}
V(\mathbf{p,p^\prime}) = V^{(-1)}+V^{(0)}+V^{(1)}+V^{(2)}+\ldots
\end{equation}
where $V^{(n)}$ is of order $v^n$ in the velocity power counting.
The first three terms in the expansion have the form
\begin{eqnarray}
&&  V^{(-1)} = {{U}_c \over {\mathbf k}^2}\, ,\nn \\
&& V^{(0)} ={ {U}_k  \over |{\mathbf k}| } \,, \\
&& V^{(1)} = U_2 + U_s\: {\bf S^2} + { U_r ({\mathbf p^2 + \mathbf 
p^{\prime 2}}) \over
  2 {\mathbf k}^2} \nn\\
  &&\qquad\quad- {i {\mathbf U}_\Lambda \cdot ({\mathbf p^\prime \times
\mathbf p})
  \over  {\mathbf k}^2 } \nn \\
&& \qquad\quad  + U_t \Big( {\mathbf \bsigma_1 \cdot \bsigma_2}-{3\,{\mathbf k
 \cdot \bsigma_1}\,  {\mathbf k \cdot \bsigma_2} \over {\mathbf k}^2} \Big )
   \,, \nn
\end{eqnarray}
where $V^{(0)} \sim 1/m$, and $V^{(1)} \sim 1/m^2$. In QCD, each of the
coefficients can be written as $U \to U^{(1)} 1 \otimes 1 +  U^{(T)} T^A
\otimes \bar T^A$, where $1$ and $T^A/\bar T^A$ are color matrices acting on
the quark/antiquark lines.  The anomalous dimensions for the coefficients
$U_c$--$U_t$ have been computed, and the details are given in
Refs.~\cite{amis,amis2,amis3}. The renormalization group improved static
potential was computed in Ref.~\cite{rgstatic}. An important point to note is
that graphs can involve both soft and ultrasoft gluons, so that the anomalous
dimensions involve \emph{both} $\alpha_s(mv)$ and $\alpha_s(mv^2)$. As an
example, the running of $U_2^{(1)}$ is given by
\begin{eqnarray}\label{u2run}
 m^2 {U}_2^{(1)}(\nu) &=& \frac{14 C_1}{3}\,
  { \alpha_s(m\nu)}{\alpha_s(m)} \ln\Big({{m\nu}\over {m}}\Big)\nonumber \\
 && -
    \frac{32\pi C_1}{3\beta_0}
    \, {\alpha_s(m)} 
 \ln\bigg[ \frac{{\alpha_s(m\nu)}}{{\alpha_s(m\nu^2)}} \bigg]
    \,
\end{eqnarray}
where $C_1=2/9$ for QCD. Note that Eq.~(\ref{u2run})
depends on $\alpha_s(m)$, $\alpha_s(m\nu)$, and $\alpha_s(m\nu^2)$. The
running coefficients in the singlet channel
($U^{(s)}=U^{(1)}-C_F U^{(T)}$) for $\bar t t$ production are  presented in
Table~\ref{tab:ttbar}.
\begin{table}
\caption{Numerical values for the $\bar tt$ singlet potentials.
The values at $\nu=1$
are the matching values at $\mu=m_t$. The values at $\nu=v$ are the velocity
renormalization group improved values, where $v=0.14$ has been used.}
\label{tab:ttbar}
\renewcommand{\arraystretch}{1.5}
\begin{center} \begin{tabular}{|c|c|c|}
\hline 
 Coefficient & 			$\nu=1$		&  $\nu=v$	\\
 \hline
${{U}_c^{(s)}}$ &		{$-1.81$}	& {$-2.47$}	\\
${{m U}_k^{(s)}}$ &		{$-0.36 $}	& {$-0.03$}	\\	
${{m^2 U}_r^{(s)}}$ & 		{$-1.81$}	& {$-1.49$}	\\
${{m^2 U}_2^{(s)}}$ &		{0} 		& {0.63} 	\\
${{m^2 U}_s^{(s)}}$ &		{0.60}		& {0.53}	\\
${{m^2 U}_\Lambda^{(s)}}$ &	{0.15}		& {0.16}	\\
${{m^2 U}_t^{(s)}}$ &		{2.71}		& {3.11} 	\\
\hline
\end{tabular} \end{center}
\end{table}
The renormalization group improved coefficients ${U}_r$ and ${U}_2$
differ significantly from their matching values, because they depend on the
ultrasoft scale through $\alpha_s(mv^2)$. The other coefficients only have a
soft anomalous dimension, and do not run as much.

\begin{figure}[t]
 \centerline{\epsfxsize=8truecm \epsfbox{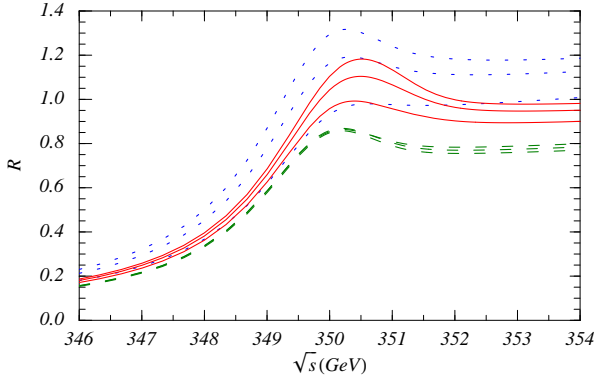}}
 \vspace{-0.5cm}
 \caption{Fixed order computation of $\bar tt$ production near
threshold. The curves are LO (dotted), NLO (dashed) and NNLO (solid.) Uses the
$1S$ mass-scheme~\cite{1Sa,1Sb,1Sc}}
\label{fig:old}
\end{figure}

\begin{figure}[t]
 \centerline{\epsfxsize=8truecm \epsfbox{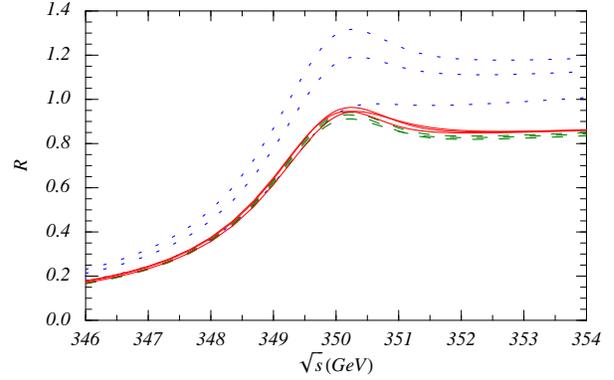}}
 \vspace{-0.5cm}
\caption{Renormalization group improved computation of $\bar tt$ production near
threshold. The curves are LL (dotted), NLL (dashed) and NNLL (solid.). Uses the
$1S$ mass-scheme~\cite{1Sa,1Sb,1Sc} }
\label{fig:new}
\end{figure}

The renormalization group improved potentials can be used to calculate the
renormalization group improved cross-section for $\bar tt$ production in the
threshold region. Fig.~\ref{fig:old} shows a sample fixed order calculation of
$R$, the ratio of $\sigma(e^+e^-\to \bar t t)/\sigma(e^+e^-\to
\mu^+\mu^-)$~\cite{review}. The scale uncertainty is of order 20\%. The
renormalization group improved version of the results is shown in
Fig.~\ref{fig:new}. There is a dramatic
reduction in the scale uncertainty, which is now around 2\%, as well as an
improvement in convergence for the normalization. The small
theoretical uncertainty means that an accurate measurement of the cross-section
can be used to study new physics. For example, a standard model Higgs boson of
mass around 115~GeV changes the cross-section by $\sim 5$\%, and is measurable.

\section{QED}

The velocity renormalization group method gives very interesting and important
results when applied to QED~\cite{amis4}. The basic potentials we will need for
QED are summarized in Table~\ref{tab:qed}. The last column gives the
contribution to the bound state energy levels due to the given potential. The
fourth column gives the order of a given potential, treating $v$ as order
$\alpha$. Since the Coulomb potential is of order unity, one finds the obvious
result that the Coulomb potential must be summed to all orders, and cannot be
treated as a perturbation. The potentials $V^{(-1)}$, $V^{(1)}$, $V^{(3)}$, are
first generated at tree-level, and are of order $\alpha$, whereas the
potentials  $V^{(0)}$, $V^{(2)}$, $V^{(4)}$, are first generated at one-loop,
and are of order $\alpha^2$.
\begin{table}[!b]
\caption{Table of potentials for QED. The first column is the potential, the
second gives typical terms in the potential, the third gives the power counting
in $\alpha$ and $v$, the fourth gives the order in the $v$ counting scheme when
$v\sim\alpha$, and the fifth gives the contribution of the potential to the
bound state energy.}
\label{tab:qed}
\renewcommand{\arraystretch}{1.5}
\begin{center}
\begin{eqnarray*}
\begin{array}{|l|c|c|c|c|}
\hline
 & & \hbox{Power Counting} & \hbox{Order} & E \\
\hline 
V^{(-1)} &  {\alpha \over \mathbf{k}^2} & {\alpha 
\over v} &  1 &  \alpha^2\\
 V^{(0)} & {\alpha \over m\vabs{\mathbf{k}}} & 
{\alpha^2 } & 
\alpha^2 & \alpha^4 \\
V^{(1)} & {\alpha \over m^2},\ {\alpha \mathbf{S}^2
\over m^2} & {\alpha  v}
 &  \alpha^2 & \alpha^4 \\
 V^{(2)} &  {\alpha \vabs{\mathbf{k}} \over m^3} 
 &  {\alpha^2 v^2} & 
  \alpha^4 &  \alpha^6 \\
 V^{(3)} & {\alpha \mathbf{k}^2 \over m^4} & 
 {\alpha v^3} &  \alpha^4 & 
  \alpha^6 \\
 \vdots &  \vdots &  \vdots &   \vdots &  \vdots\\
 \hline
\end{array}
\end{eqnarray*}
\end{center}
\end{table}

The bound state energy levels can be determined to order $\alpha^4$ by
computing the matrix elements of $V^{(0)}$ and $V^{(1)}$ between Coulomb
wavefunctions. Time-ordered products of two potentials, such as $T\left[ 
V^{(0)} V^{(0)} \right]$, $T\left[  V^{(0)} V^{(1)} \right]$ and $T\left[ 
V^{(1)} V^{(1)}  \right]$ first contribute at order $\alpha^6$. In principle,
to obtain the energy levels to order $\alpha^4$, one also needs the one- and
two-loop matching corrections to the Coulomb potential. However, such
corrections vanish in QED. As a result, the first correction to the order
$\alpha^2$ binding energy is of order $\alpha^4$, and is given by the matrix
element of $V^{(0)}+V^{(1)}$. There are no order $\alpha^3$ corrections to the
energy levels in QED.

Define the leading and next-to-leading order anomalous dimensions of a
potential to be the anomalous dimension from graphs at one and two higher
orders in $\alpha$ than the potential itself. For $V^{(-1)}$ and  $V^{(1)}$
which are of order $\alpha$, the leading order anomalous dimension is of order
$\alpha^2$, and the next-to-leading order anomalous dimension is of order
$\alpha^3$. For $V^{(0)}$ which is of order $\alpha^2$, the leading order
anomalous dimension is of order $\alpha^3$, and the next-to-leading order
anomalous dimension is of order $\alpha^4$. Since different terms in the
potential are of different orders in $\alpha$, the terms leading and
next-to-leading order are not related to the number of loops.

Integrating the renormalization group equations for  $V^{(0)}$ and $V^{(1)}$
using the leading order anomalous dimension gives a series of the form
\[
\alpha\left (1+ \alpha\ln\alpha + \alpha^2 \ln^2 \alpha +  \alpha^3 \ln^3
\alpha + \ldots \right),
\]
which contributes
\[ \label{LOenergy}
\alpha^4
 \left(1+ \alpha\ln\alpha + \alpha^2
\ln^2 \alpha +  \alpha^3 \ln^3 \alpha + \ldots \right)
\]
to the energy. Integrating the next-to-leading order anomalous dimensions
gives 
\[ \label{NLOenergy}
\alpha^4 \alpha
 \left(1+ \alpha\ln\alpha + \alpha^2
\ln^2 \alpha +  \alpha^3 \ln^3 \alpha + \ldots \right)
\]
terms in the energy. The next-to-next-to-leading anomalous dimension
gives
\[
\alpha^4 \alpha^2 \left(1+ \alpha\ln\alpha + \alpha^2
\ln^2 \alpha +  \alpha^3 \ln^3 \alpha + \ldots \right),
\]
terms in the energy, which are the same order as those obtained by using the
leading order anomalous dimension for the $V^{(2)}$ and $V^{(3)}$ potentials
which first contribute at order $\alpha^6$. Thus one can compute the 
\[
\begin{array}{llll}
\alpha^5 \ln \alpha & \alpha^6 \ln^2 \alpha & \alpha^7 \ln^3 \alpha & \ldots\\
\alpha^6 \ln \alpha & \alpha^7 \ln^2 \alpha & \alpha^8 \ln^3 \alpha & \ldots\\
\end{array}
\]
series in the energy 
using $\gamma_{\rm LO}$, $\gamma_{\rm NLO}$ for $V^{(0,1)}$.

\section{LEADING ORDER}

\begin{figure}
  \centerline{\epsfxsize=1.75truecm\epsfbox{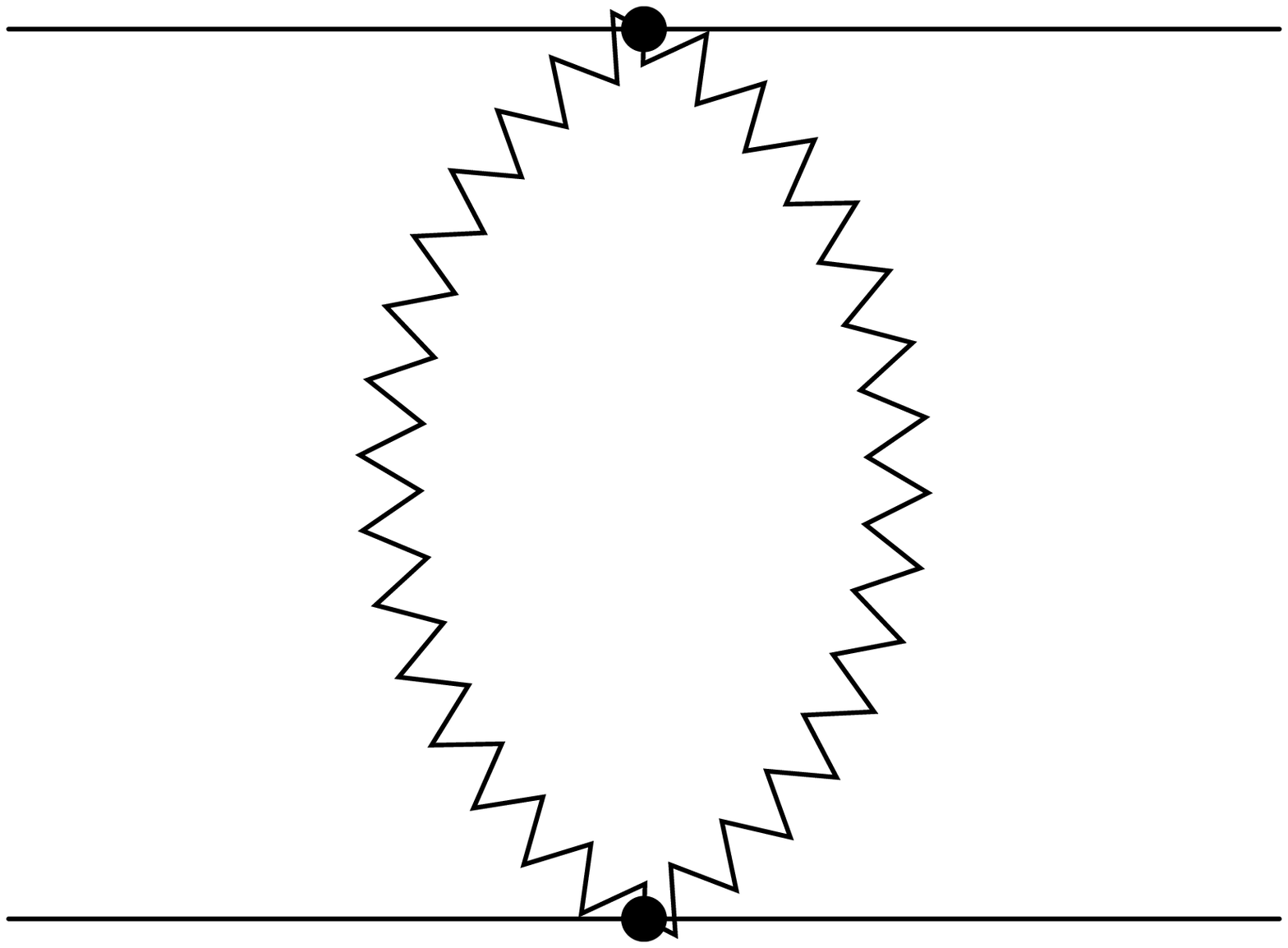}
  \quad\epsfxsize=1.75truecm\epsfbox{0027_fd10.eps}
  \quad\epsfxsize=1.75truecm\epsfbox{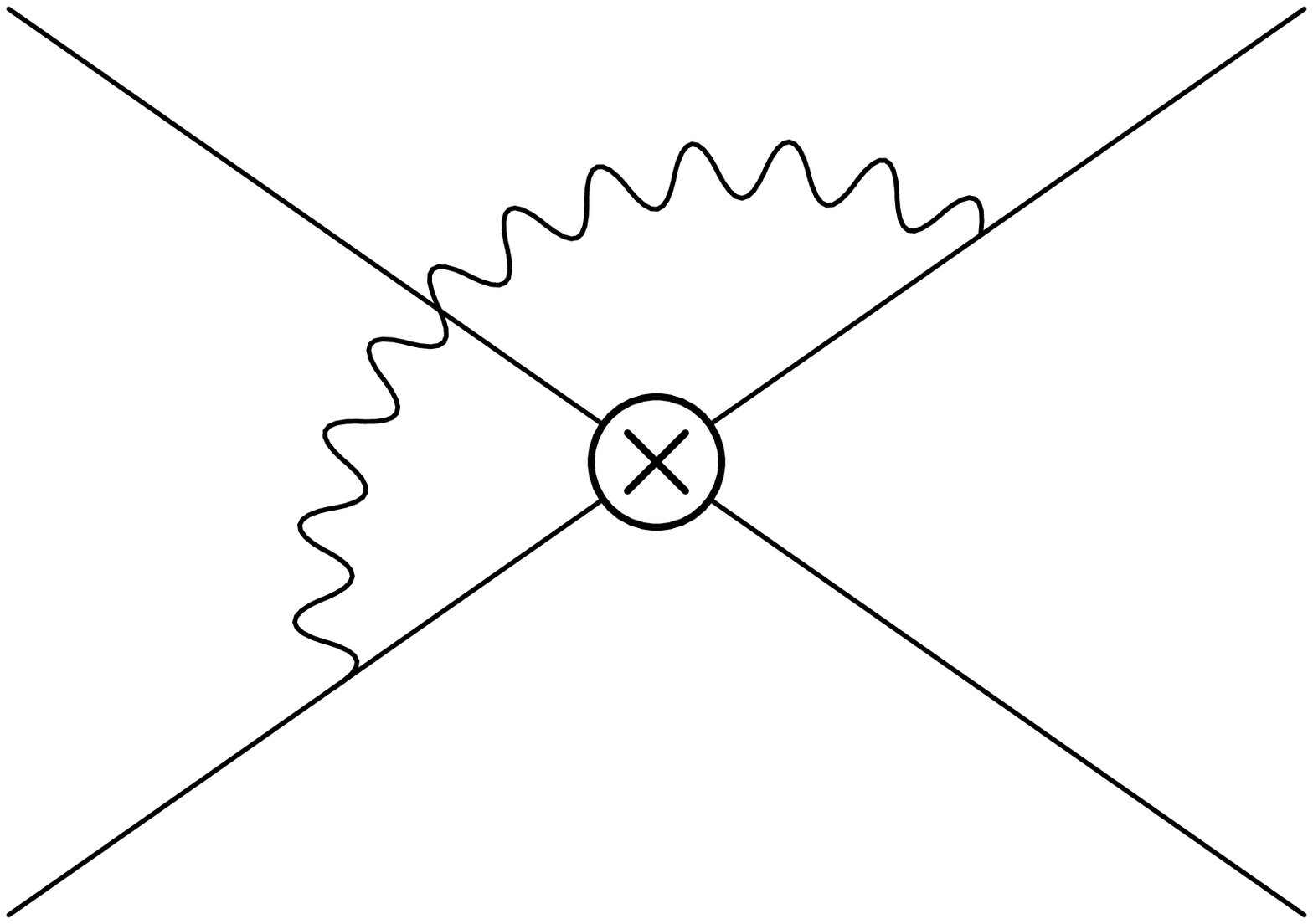}}
\caption{One-loop running of $V^{(1)}$ in QED. The first graph has a soft
photon, and the other graphs have ultrasoft photons and a potential.}
\label{fig:23}
\end{figure}

The Coulomb potential and $V^{(0)}$ do not run in QED at leading and
next-to-leading order, so one is left with the running of $V^{(1)}$. The
anomalous dimensions are evaluated for a particle of mass $m_1$ and charge $-e$
interacting with a second particle of mass $m_2$ and charge $Ze$. Evaluating
the graphs in Fig.~\ref{fig:23} gives
\begin{equation}
\nu {d U_2 \over d \nu} = {{14 Z^2 \alpha^2 \over 3 m_1 m_2}}+
{{2}\alpha \over 3 \pi}\left({1\over m_1}
   + {Z \over  m_2}\right)^2 U_c
\end{equation}
where the first term is the soft contribution from Fig.~\ref{fig:23}a and the
second is the ultrasoft contribution from Fig.~\ref{fig:23}b,c. Note that the
ultrasoft contribution has been multiplied by two, since the anomalous dimension
for the velocity renormalization group is $\gamma_S + 2 \gamma_U$. The other
coefficients in $V^{(1)}$ ($U_r$, $U_s$, $U_\Lambda$, $U_t$) have zero
anomalous dimension at this order.

Since the Coulomb potential and $\alpha$ do not run in QED, one can combine the
two terms,
\begin{equation}
\label{20}
\nu {d U_2 \over d \nu} = \gamma_0 U_c
\end{equation}
which defines
\begin{equation}
\gamma_0 = {2 \alpha \over 3 \pi }\left( {1\over m_1^2} + {Z \over 4
    m_1 m_2} + {Z^2 \over m_2^2} \right).
\end{equation}
$\gamma_0$ is a constant in QED since $\alpha$ does not run. Integrating
Eq.~(\ref{20}) gives
\begin{equation}
\label{lo}
U_2(\nu) = U_2(1) + \gamma_0 U_c \ln \nu,
\end{equation}
where $U_2$ is evaluated at $\nu=v=\alpha$. Since $\gamma_0$ is a constant,
$U_2(\nu)$ only has a $\ln\nu$ term, and terms of the form $\ln^n \nu$, with
$n>1$ vanish. As a result, the leading order energy series Eq.~(\ref{LOenergy})
terminates after a single term, so one has an $\alpha^5 \ln \alpha$
contribution to the energy, but the $\alpha^6 \ln^2 \alpha$, etc.\ terms
vanish. At low orders, the absence of terms other than $\alpha^5\ln \alpha$ in
the leading order series has been noticed before by an explicit examination of
Feynman graphs. This is the first general proof that all the terms beyond 
$\alpha^5\ln \alpha$ in the leading order series vanish

The matrix element of $U_2$ gives the energy shift
\begin{eqnarray}
\Delta E &= & \vev{U_2(\nu)} \nn\\
&=&{\gamma_0} U_c \ln \nu \abs{\psi(0)}^2 \\
&=& -{8 Z^4 \alpha^5 m_R^3 \over 3 \pi n^3}\left( {1\over m_1^2} +
 {Z \over 4 m_1 m_2} + {Z^2 \over m_2^2} \right)  \ln Z \alpha ,\nn
\end{eqnarray}
where we have used
\begin{eqnarray}
 \abs{\psi(0)}^2 = {(m_R Z \alpha)^3 \over \pi n^3}
\end{eqnarray}
for the $nS$ state, and $m_R$ is the reduced mass. This is the famous
$\alpha^5 \ln \alpha$ correction to the Lamb shift first computed by Bethe,
including all recoil corrections.

\section{NEXT-TO-LEADING ORDER}

At next-to-leading order, the anomalous dimension for $V^{(1)}$ is
\begin{eqnarray}
 &&  \left. \nu {d U_{2+s} \over d\nu}  \right|_{\rm NLO} =
   \rho_{ccc}\, U_c^3 + \rho_{cc2}\, U_c^2  \left({ U_{2+s}}
+ U_r
   \right) \nonumber \\
 &&\quad + {\rho_{c22}}\, U_c\left({U_{2+s}^2}
 +2{U_{2+s}} U_r + \frac34 U_r^2
   -5 U_t^2 {\bf S^2} \right) \nonumber \\
 && \quad +\rho_{ck}\, U_c U_k +\rho_{k2}\, U_k \left({U_{2+s}}  +
   U_r/2\right) \nonumber \\
 && \quad + \rho_{c3}\, U_c \left({U_3}+U_{3s} S^2 + {1\over2}U_{rk}\right)
  , \label{nlo}
\end{eqnarray}
where $U_{2+s}=U_2 + U_s {\mathbf S}^2$, and the coefficients are
\begin{eqnarray}
 \rho_{ccc}&=& -{m_R^4 \over 64 \pi^2} \left( {1\over m_1^3} + {1\over
   m_2^3}\right)^{\!2},\nn \\
 \rho_{c22}&=& -{m_R^2 \over 4 \pi^2},\nn \\
 \rho_{cc2}&=& -{m_R^3 \over 8 \pi^2} \left( {1\over m_1^3} +
   {1\over m_2^3} \right),\nn \\
 \rho_{c3} &=& { 2 m_R \over  \pi^2},\nn \\
 \rho_{ck} &=& {m_R^2 \over 2 \pi^2}\left({1\over m_1^3} +{1\over m_2^3}
    \right), \nn \\
 \rho_{k2} &=& { 2 m_R \over  \pi^2}.
\end{eqnarray}

The anomalous dimension Eq.~(\ref{nlo}) can be integrated by substituting the
leading order running, Eq.~(\ref{lo}) for the coefficients on the right-hand
side. Since only $U_2$ runs at leading order, the right hand side has at most a
$\ln^2 \nu$, so that the integral has at most a $\ln^3 \nu$ term. This implies
that the next-to-leading order series Eq.~(\ref{NLOenergy}) terminates after the
first three terms, $\alpha^6\ln \alpha$,  $\alpha^7\ln^2 \alpha$, and
$\alpha^8\ln^3 \alpha$.

\subsection{$\ln^3 \alpha$}

The only term that contributes to the $\ln^3\alpha$ correction is the $U_2^2$
term of Eq.~(\ref{nlo}). Integrating gives a contribution to $U_2(\nu)$ of the
form
\begin{eqnarray} \label{a3}
{1\over 3}\, {\gamma_0^2}\, 
{\rho_{c22}}\, {U}_c^3(1)\, \ln^3\nu,
\end{eqnarray}
which is spin-independent, and has no imaginary part. There is no contribution
to the decay width or hyperfine splitting at this order. The Lamb shift at this
order is obtained by multiplying Eq.~(\ref{a3}) by the matrix element of the
unit operator, $\abs{\psi(0)}^2$, to give
\begin{eqnarray}
\Delta E &=& {64 m_R^5 \alpha^8 Z^6 \over 27 \pi^2 n^3}
\ln^3 \left(Z \alpha \right)\nn\\
&& \times \left(
{1\over m_1^2} + {{Z\over 4 m_1 m_2} + {Z^2\over  m_2^2}}
 \right)^2
\end{eqnarray}
which is approximately 8~KHz for the $2P$--$2S$ Lamb shift in Hydrogen.
Substituting $Z=1$ and $m_1=m_2=m_e$ gives the $\alpha^8\ln^3\alpha$ Lamb shift
for positronium
\begin{eqnarray}
\Delta E &=&
{3 m_e \alpha^8 \ln^3 \alpha \over 8 \pi^2 n^3}.
\end{eqnarray}
The positronium Lamb shift is a new result, as are the recoil terms in the
Hydrogen Lamb shift. In the limit $m_1/m_2 \to 0$, the Hydrogen Lamb shift has
been computed previously by several groups. There is an analytic computation by
Karshenboim~\cite{karshenboim} and a numerical computation by Goidenko et
al.~\cite{goidenko} that agree with our result. There are also numerical
computations by Malampalli and Sapirstein~\cite{malampalli}, and by
Yerokhin~\cite{yerokhin} which agree with each other, but disagree with the
other results. Recently, there has been a computation by
Pachucki~\cite{pachuckinew} that agrees with our result.
Yerokhin~\cite{yerokhinnew} has emphasized that the complete $\alpha^8 \ln^3
\alpha$ Lamb shift might not be contained in the loop-after-loop calculations
of Refs.~\cite{malampalli,yerokhin}.

\begin{figure}
  \centerline{\epsfxsize=4truecm\epsfbox{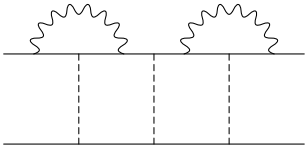}}
\caption{Four-loop diagram that contributes to the $\alpha^8 \ln^2 \alpha$ Lamb
shift}
\label{fig:21}
\end{figure}

The other calculations rely on extracting the logarithm from four-loop diagrams
such as Fig.~\ref{fig:21}. The velocity renormalization group factors the graph
into the product of a two-loop anomalous dimension $\rho_{c22}$, and the square
of a one-loop anomalous dimension $\gamma_0^2$.

\subsection{$\ln^2 \alpha$}

The $\ln^2 \alpha$ contribution to the hyperfine splitting and decay widths is
given by the $U_2(\nu)$ contribution
\begin{eqnarray}
 {\gamma_0}\, {\rho_{c22}}\, {U}_c^2(1) \left[{U}_2(1)+ {U}_s(1) {\mathbf S}^2
  \right] \ln^2 \nu + \ldots.
\end{eqnarray}
The spin-dependent term gives the $\alpha^7 \ln^2 \alpha$ hyperfine splitting
\begin{eqnarray}
\hbox{HFS} &=&  -{64 Z^6 \alpha^7 m_R^5 \mu_1 \mu_2 \over 9m_1m_2 \pi n^3} 
  \ln^2 (Z \alpha)\nn\\
  &&\times \left[{1\over m_1^2} +  {Z\over 4 m_1 m_2} +  {Z^2\over  m_2^2} 
  \right]
\end{eqnarray}
where $\mu_i$ are the magnetic moments normalized to unity for a Dirac fermion.
Our result agrees with previous calculations~\cite{karshenboim,thesis}.
Substituting the matching values for $U_s(1)$ for positronium (which differs
from Hydrogen because of annihilation contributions), one finds the positronium
hyperfine splitting
\begin{eqnarray}
\hbox{Ps HFS}=  -{7  m_e \over 8 \pi n^3}\ \alpha^7 \ln^2\! \alpha ,
\end{eqnarray}
which agrees with a recent computation of Melnikov and
Yelkhovsky~\cite{melnikov}. The imaginary parts of the matching coefficients
give the decay widths~\cite{karshenboim},
\begin{eqnarray}
 {\Delta \Gamma\over \Gamma_0} = \gamma_0\, \rho_{c22}\, {U}_c(1)^2 \ln^2 \nu =
  -{3  \over 2 \pi}\alpha^3 \ln^2\! \alpha,
\end{eqnarray}
for both ortho- and para-positronium.

\subsection{$\ln \alpha$}
The $\ln \alpha$ contributions to the decay width arise from
\begin{eqnarray}
&& U_{2+s} \Big[ {\rho_{c22}} U_c \left( {U}_{2+s} +
2  U_r \right) \nn\\
&&\qquad + \rho_{cc2}
U_c^2 + \rho_{2k} U_k \Big] \ln \nu +\ldots.
\end{eqnarray}
which give
\begin{eqnarray}
{\Delta \Gamma \over \Gamma_0} &=& \left({m_e^2 \over 2 \pi} \hbox{Re}\, U_{2+s}
 - 2 \right)\ln \nu \nn \\
  &=& \left( {7 {\mathbf S}^2 \over 6} - 2 \right)
 {\alpha^2} \ln \alpha,
\end{eqnarray}
so that
\begin{eqnarray}
\left({\Delta \Gamma \over \Gamma_0}\right)_{\rm ortho} &=&
  {\alpha^2 \over 3}\ln \alpha \,, \nn \\
\left({\Delta \Gamma \over \Gamma_0}\right)_{\rm para} &=&
  -2{\alpha^2}\ln \alpha \,.
\end{eqnarray}
These agree with existing results~\cite{caswell2,khriplovich}.

\section{CONCLUSIONS}

The methods presented here give a systematic way of separating scales in
nonrelativistic bound state problems. All large logarithms are summed using the
velocity renormalization group. The method provides a universal description of
QED logarithms. The agreement with known results at order $\alpha^5 \ln
\alpha$, $\alpha^6 \ln \alpha$, $\alpha^7 \ln^2 \alpha$, and $\alpha^8 \ln^3
\alpha$ is a highly non-trivial check of the formalism. In QED, one finds that
the leading order series terminates after one term, and the next-to-leading
order series terminates after three terms. In addition, the method resolves a
controversy about the $\alpha^8 \ln^3 \alpha$ Lamb shift for Hydrogen, and
gives the first calculation of the $\alpha^8 \ln^3 \alpha$ energy shift for
positronium.

In QCD, one can distinguish $\alpha_s(mv)$ and $\alpha_s(mv^2)$, and both can
appear simultaneously in the same anomalous dimension. The renormalization
group improved potentials can be used to compute $\bar tt $ production, and
reduce the scale uncertainties by a factor of ten.

The velocity renormalization group should also be applicable to other problems
with correlated scales. In the bound state problem, one can generate the scale
$mv$ in loop graphs from the scale $m$ and $mv^2$, $mv=\sqrt{m \times mv^2}$.
Similar effects can occur at finite temperature, where one has the scales $T$,
$gT$ and $g^2T$, and some of the ideas described here might be applicable to
that problem as well.

This work was supported in part by the Department of Energy under grant
DOE-FG03-97ER40546 and by NSERC of Canada.

\end{document}